\documentclass{sig-alternate-05-2015}
\usepackage{graphicx}

\begin{document}

\setcopyright{acmcopyright}





%

\title{Combating Malicious DNS Tunnel}
%
%
%
%
%

\numberofauthors{1} 
%
\author{
%
%
Zheng~Wang\\
       \affaddr{National Institute of Standards and Technology, USA}\\
       \email{zhengwang98@gmail.com}
}

\maketitle

\section{Introduction}
The Domain Name System (DNS) is a fundamental Internet infrastructure, which resolves billions of queries per day in support of global communications and commerce. The most common use of DNS is to map human-friendly domain names to machine-readable IP addresses.The DNS is designed based on the client-server model where stub resolver at the client side originates DNS query for some query name and authoritative server at the server side responses with the requested mapping associated with the query name. To simplify client and enhance the scalability and efficiency of name resolution, stub resolver commonly relies on recursive resolver to traverse the DNS tree and return final answer. The DNS is known to be susceptible to cache poisoning and man-in-the-middle attacks, so the major efforts devoted to securing DNS in the past two decades focused on ensuring source authentication and data integrity (such as the DNSSEC initiatives), which are essential for the common use of DNS. As the DNS is taken for granted an indispensable service for almost every Internet end user, it is also convenient for malicious use. An emerging misuse of DNS in recent years is DNS tunnel. Unlike the common use of DNS which aims at finding the mapping data associated with the interested query name, the goal of DNS tunnel is to use DNS as a communication stack between the querier and the responder. A DNS tunnel can be used for ``command and control'', data exfiltration or tunneling of any internet protocol (IP) traffic.

There are a variety of services that leverage DNS tunnel to convey specific information about their users to their providers.For example, Sophos \cite{Sophos} designs and maintains a protocol/framework to encode generic information about the threat and the detection, which is based on DNS transaction. When a Sophos-enabled endpoint triggers a detection by a scanner and needs to look up the security services, it requests the sophosxl.net name servers using a specially crafted DNS query. The domain in the DNS query is generated to include all necessary information about the suspicious file. Then the endpoint adjust its behavior according to the information encoded in the DNS response. Some other anti-virus software vendors also have similar services or systems. McAfee \cite{McAfee} provides a reputation system through the DNS channel. When a suspicious file is found by the McAfee software, a query can be generated with a fingerprint and other information and sent to the McAfee server. And the McAfee server encodes the reputation in the response. The response could be an address from 127.0.0.0/16 indicating a specific code of reputation or NXDOMAIN indicating that the file is not known to hold malicious content. Those uses of DNS tunnel are generally considered as benign since they basically intend to serve the users rather than jeopardize them.

Recent years also witness the increasingly prevalent malicious use of DNS tunnel. In a 2012 presentation at the RSA conference, Ed Skoudis \cite{P1} identified that DNS based Command and Control of malware as one of the six most dangerous new attacks. Ed shared that ``Attackers have recently used this technique in cases involving the theft of millions of accounts''. It has been shown that DNS tunneling can achieve bandwidth of 110 KB/s with latency of 150 ms \cite{P2}.A recent example of malicious use of DNS tunnel is a new variant of a point of sale (POS) malware family \cite{POS} which exfiltrates stolen payment card data over DNS. That malware, once executed on the targeted host, will collect sensitive information with user information and card data included, encode them with a custom Base32 encoding algorithm, and then makes a DNS query with this information to a hardcoded domain. The attacker will be thus by notified with data which typically is sufficient in most scenarios to attempt card fraud.

Given the growing threats posed by malicious use of DNS tunnel, the defense is still unfortunately scarce by now. The reason is basically of two fold. One is caused by the fact that DNS traffic is often not monitored, restricted nor blocked unless it potentially amounts to a DoS or DDoS like devastating level. It is widely believed that DNS should be universally available to anyone who would like to use it. Therefore when malicious communication on e.g., HTTP or FTP by attackers are subject to defensive monitoring and blocking especially in sensitive environments, DNS tunnel is often unlikely to be disabled by blocking. The other is caused by the fact that benign and malicious use of DNS tunnel co-exist and no effective way of differentiating them is available. Once a DNS tunnel is detected, simply blocking it risks disabling a benign use of it and simply allowing it risks enabling a malicious use of it.

This paper proposes a defense scheme against malicious use of DNS tunnel. A tunnel validator is designed to provide trustworthy tunnel-aware defensive recursive service. In addition to the detection algorithm of malicious tunnel domains, the tunnel validation relies on registered tunnel domains as whitelist and identified malicious tunnel domains as blacklist. A benign tunnel user is thus motivated to register its tunnel domain before using it. Through the tunnel validation, the secure domains are allowed to the recursive service provided by the tunnel validator and the insecure domains are blocked. All inbound suspicious DNS queries are recorded and stored for forensics and future malicious tunnel detection by the tunnel validator.

\begin{figure}[!t]
\centering
\includegraphics[width=3.2in]{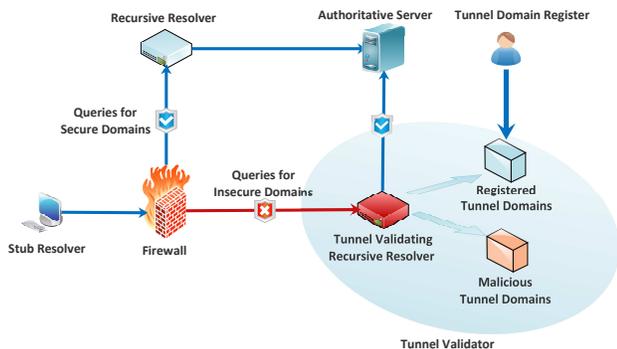}
\caption{The defense scheme against malicious use of DNS tunnel.}
\end{figure}

\section{The Defense Scheme}

\subsection{The User End}

A stub resolver installed on the host originates a DNS query for a tunnel domain. Note that the host, if compromised by the attacker, is unlikely to perform any tunnel detection or filtering before a malicious insecure domain is requested. So the suspicious insecure domains are more likely to be identified by the firewall. The tunnel detection algorithm implemented by the firewall relies on the payload analysis of the DNS message with the following techniques:

$\bullet$ \textbf{Query name length}. Compared with other alternative protocols, DNS message is very limited in size for tunnel utilities.For example, to make full use of DNS message, an attacker would place as much as possible data into an encoded query name to maximum the data transfer rate over DNS. So a tunnel query name has a good chance to have long labels of up to 63 characters and long overall length of up to 255 characters.

$\bullet$ \textbf{Query name entropy}. Common use of DNS often has query name with dictionary words or something that looks meaningful.By contrast,tunnel query names generally have a higher entropy and a more even use of the character set. More accurately, a set of features such as the maximum, minimum, average, median, and variance of the entropy of labels within a query name can be calculated as an indicator of tunnel query names.

$\bullet$ \textbf{Statistical analysis}. The statistical characteristics of the characters within a query name provides another means of detect tunneling.Common query names tend to have few numbers whereas tunnel query names may have more numbers.Other statistical characteristics include the percentage of the length of the Longest Meaningful Substring (LMS), the number of unique characters, etc.

To further detect the suspicious insecure domains from the suspicious tunnel domains, we need to employ the response features. Most malicious use of DNS tunnel care little about responding, because all necessary information needed by attackers are completely encoded in the tunnel domains unidirectionally delivered from the users to the attackers. To simplify the responding to diversely distributed query names, the attackers often provide uniform responses to the tunnel domains. The uniform responses include NXDOMAIN, NODATA, constant answers featured by wildcard responses, SERVFAIL, and even no response at all. So if the suspicious tunnel domains below a domain are always uniformly responded, that domain has a good chance of being malicious tunnel domain and thus may be identified as suspicious insecure domain.

Based on the detection of suspicious insecure domains, the firewall forwards the queries for suspicious insecure domains to the tunnel validating recursive resolver and the remaining queries to the normal recursive resolver.

\subsection{The Tunnel Validator}

The tunnel validator is designed to identify the insecure query domains and block their queries and at the same time provide recursive service to the queries for secure query domains.

If identified as secure, the query domain has its query served by the tunnel validating recursive resolver. That is, the queries hitting the cache may be responded immediately from the cache, and only the cache missed queries have to be forwarded to the authoritative server. If identified as insecure, the query domain may its query blocked by the tunnel validating recursive resolver. The possible blocking policy may include a NXDOMAIN or SERVFAIL response or no response at all.

As the core function of the tunnel validator, the classification between malicious tunnel domains and benign tunnel domains is based on three mechanisms:

$\bullet$ \textbf{Tunnel domain registration}. The service provider, who desires to use DNS tunnel by generating encoded domains and issuing queries for them, should register its intended domain for DNS tunnel and the pattern of generating tunnel domains with the DNS tunnel registry. The DNS tunnel registry may be hosted by the tunnel validator or the third party which allows access to the shared registration system (SRS) from the tunnel validator. When handling the inbound queries, the tunnel validator should look up the query domains from the SRS of the DNS tunnel registry. Any query domain which hits an entry of the SRS and satisfies its pattern defined by the DNS tunnel register should be classified as secure. So the registered tunnel domains function as whitelist.

$\bullet$ \textbf{Malicious domain registration}. The malicious domains identified by the outside parties may be submitted to and used by the tunnel validator. Another source of known malicious domains is from the history of classification performed by the tunnel validator itself. Any query domain falling into the maintained malicious domains should be classified as insecure. So the registered tunnel domains function as blacklist.

$\bullet$ \textbf{Detection algorithm of malicious tunnel domain}. For those domains not covered by whitelisting and blacklisting, the tunnel validator implements detection algorithm of malicious tunnel domain. The detection algorithm may apply the payload analysis of the DNS message as well as the inspection of responses. The techniques used by the detection algorithm may be similar to those by the firewall discussed above. Moreover, the caching performed by the tunnel validating recursive resolver enables another detection method against the malicious use of DNS tunnel. If the tunnel validator is heavily and widely relied on by the user end systems, an abundant amount of secure domains and insecure domains will be possessed and cached by the tunnel validating recursive resolver. Given that secure domains are likely to have a high cache hit rate and insecure domain are likely to have a low cache hit rate (largely because they are mostly used once), a domain hitting cache should be rated with a better chance of a secure domain.

The tunnel validator should record and keep the log files of the inbound and outbound DNS queries or at least those of the outbound DNS queries. In case of false negatives, the outbound queries containing malicious tunnel domains are falsely forwarded by the tunnel validating recursive resolver to the authoritative servers in control of the attackers. Thus the log files of the outbound DNS queries may facilitate forensics against the malicious use of DNS tunnel.

\subsection{The Tunnel Domain Register}
The benign DNS tunnel users are highly motivated to register their intended tunnel domains because those domains may be otherwise blocked by the tunnel validator. For those malicious DNS tunnel users who may also attempt to register their intended malicious tunnel domains, they are both regulated by the agreement of tunnel registration and restricted by the dedicated forensics enabled by logging. While the encoded information in the malicious tunnel domains are usually stealthy against monitoring and censoring, the source and the target of tunneled information are hardly stealthy. So the attackers residing at the authoritative side, if misbehaves using their registered tunnel domains, can be traced from the logging and addressed technically and legally. One technical response to a confirmed malicious registered domain is to invalidate it from the SRS and add it to the blacklist of malicious domains.

%
\bibliographystyle{abbrv}
\bibliography{T00}  
\end{document}